\documentclass[twocolumn, trackchanges]{aastex6}
\shortauthors{Plotkin et al.}

\begin{document}

\title{Up and Down the Black Hole Radio/X-ray Correlation: the 2017 mini-outbursts from Swift J1753.5$-$0127}
\shorttitle{mini-outbursts from Swift J1753.5$-$0127}

\author{
R.~M.~Plotkin,\altaffilmark{1}
J. Bright,\altaffilmark{2}
J.~C.~A.~Miller-Jones,\altaffilmark{1}
A.~W.~Shaw,\altaffilmark{3}
J.~A.~Tomsick,\altaffilmark{4}
T.~D.~Russell,\altaffilmark{5}
G.-B.Zhang,\altaffilmark{6}
D.~M.~Russell,\altaffilmark{6} 
R.~P.~Fender,\altaffilmark{2}
J.~Homan,\altaffilmark{7,8}
P.~Atri,\altaffilmark{1}
F.~Bernardini,\altaffilmark{6}
J.~D.~Gelfand,\altaffilmark{6}
F.~Lewis,\altaffilmark{9,10}
T.~M.~Cantwell,\altaffilmark{11}      
S.~H.~Carey,\altaffilmark{12}           
K.~J.~B.~Grainge,\altaffilmark{11}   
J.~Hickish,\altaffilmark{13,12}           
Y.~C.~Perrott,\altaffilmark{12}         
N.~Razavi-Ghods,\altaffilmark{12}  
A.~M.~M.~Scaife,\altaffilmark{11}    
P.~F.~Scott,\altaffilmark{12}             
D.~J.~Titterington.\altaffilmark{12}   
}
\altaffiltext{1}{International Centre for Radio Astronomy Research - Curtin University, GPO Box U1987, Perth, WA 6845, Australia; richard.plotkin@curtin.edu.au}
\altaffiltext{2}{Department of Physics, Astrophysics, University of Oxford, Keble Road, Oxford OX1 3RH, UK}
\altaffiltext{3}{Department of Physics, University of Alberta, 4-181 CCIS, Edmonton, AB T6G 2E1, Canada}
\altaffiltext{4}{Space Sciences Laboratory, 7 Gauss Way, University of California, Berkeley, CA 94720-7450, USA}
\altaffiltext{5}{Anton Pannekoek Institute for Astronomy, University of Amsterdam, Science Park 904, NL-1098 XH Amsterdam, the Netherlands}
\altaffiltext{6}{New York University Abu Dhabi, PO Box 129188, Abu Dhabi, UAE}
\altaffiltext{7}{MIT Kavli Institute for Astrophysics and Space Research, 70 Vassar Street 37-582D, Cambridge, MA 02139, USA}
\altaffiltext{8}{SRON, Netherlands Institute for Space Research, Sorbonnelaan 2, 3584 CA Utrecht, The Netherlands}
\altaffiltext{9}{Faulkes Telescope Project, School of Physics \& Astronomy, Cardiff University, The Parade, CF24 3AA, Cardiff, Wales}
\altaffiltext{10}{Astrophysics Research Institute, Liverpool John Moores University, 146 Brownlow Hill, Liverpool L3 5RF, UK}
\altaffiltext{11}{Jodrell Bank Centre for Astrophysics, Alan Turing Building, School of Physics and Astronomy, University of Manchester, M13 9PL, UK}
\altaffiltext{12}{Astrophysics Group, Cavendish Laboratory, 19 J. J. Thomson Avenue, Cambridge CB3 0HE}
\altaffiltext{13}{Radio Astronomy Laboratory, University of California, Berkeley, CA 94720, USA}

\newcommand{\fullsrcname}{Swift J1753.5-0127}
\newcommand{\src}{J1753}
\newcommand{\hsevfull}{H1743$-$322}
\newcommand{\hsev}{H1743$-$322}
\newcommand{\gxthree}{GX 339$-$4}
\newcommand{\vfour}{V404~Cygni}
\newcommand{\xtejeleven}{XTE~J1118+480}
\newcommand{\xtejfifteen}{XTE~J1550$-$564}
\newcommand{\xtejseventeen}{XTE~J1752$-$223}
\newcommand{\maxijsixteen}{MAXI~J1659$-$152}
\newcommand{\asix}{A~0620$-$00}
\newcommand{\mwcsix}{MWC~656}
\newcommand{\xrb}{BHXB}

\newcommand{\lr}{L_{\rm R}}
\newcommand{\lx}{L_{\rm X}}
\newcommand{\ledd}{L_{\rm Edd}}
\newcommand{\flux}{{\rm erg\,s^{-1}\,cm^{-2}}}
\newcommand{\ergs}{{\rm erg\,s^{-1}}}
\newcommand{\nh}{N_{\rm H}}
\newcommand{\cmtwo}{{\rm cm}^{-2}}
\newcommand{\msun}{M_\odot}
\newcommand{\mbh}{M_{\rm BH}}
\newcommand{\Mdot}{\dot{M}}
\newcommand{\mdot}{\dot{m}}

\newcommand{\note}[1]{\authorcomment1{#1}}

\begin{abstract}
The candidate black hole X-ray binary  \fullsrcname\  faded  to quiescence in 2016 November,  after a prolonged outburst that was discovered in 2005.  Nearly three months later the system displayed renewed activity that lasted through 2017 July.  Here, we present radio and X-ray monitoring over $\approx$3 months of the renewed activity to study the coupling between the  jet and the inner regions of the disk/jet system.   Our observations cover low X-ray luminosities that have not historically  been well-sampled ($\lx \approx 2\times10^{33} - 10^{36}~\ergs$; 1-10 keV), including time periods when the system was both brightening and fading.  At these low luminosities \fullsrcname\ occupies a parameter space in the radio/X-ray luminosity plane that is comparable to ``canonical'' systems (e.g., GX 339$-$4),  regardless of whether  the system was brightening or fading, even though during its  $\gtrsim$11-year outburst \fullsrcname\ emitted less radio emission from its  jet than expected.   We discuss implications for the existence of a single radio/X-ray luminosity correlation for black hole X-ray binaries at the lowest luminosities ($\lx \lesssim 10^{35}~\ergs$), and we compare to supermassive black holes.  Our campaign  includes the lowest luminosity quasi-simultaneous radio/X-ray detection to date for a black hole X-ray binary during its rise out of quiescence, thanks to early notification from optical monitoring combined with fast responses from sensitive multiwavelength facilities. 
\end{abstract}



 \section{Introduction}
\label{sec:intro}

Black holes in X-ray binary systems (\xrb s) spend the majority of their time accreting relatively weakly, in a regime where a non-negligible  fraction of  their accretion power is channeled into  compact relativistic jets  \citep{fender03, kording06}.   We define weakly accreting systems here as \xrb s with X-ray luminosities $\lx \lesssim 10^{37}~\ergs$, or similarly, Eddington ratios $\lx/\ledd \lesssim 0.01$,\footnote{The Eddington luminosity $\ledd=1.3\times10^{38} \left( \mbh/\msun\right)~\ergs$, which we approximate as $\ledd \sim 10^{39}~\ergs$ here for a $\approx$10~$\msun$ black hole.} 
which covers both the ``hard'' X-ray spectral state ($10^{-5} \lesssim \lx/\ledd \lesssim 10^{-2}$; \citealt{remillard06}) and ``quiescence'' ($\lx/\ledd \lesssim 10^{-5}$; \citealt{plotkin13}).  
 When weakly accreting \xrb s change their luminosities over day-to-week timescales, they trace out distinct paths through the radio luminosity ($\lr$) -- X-ray luminosity ($\lx$) plane \citep[e.g.,][]{corbel13, gallo14}.   The radio emission  is partially self-absorbed synchrotron radiation from  a steady, unresolved, flat-spectrum  jet \citep{blandford79, fender01}, while the X-rays probe the inner regions of the accretion flow/jet.  Thus, the presence of correlated radio and X-ray variability suggests a physical connection between the  jet and the emission regions closest to the black hole \citep{heinz03, markoff03}.  

Three \xrb\ systems (\gxthree, \vfour, and \xtejeleven) are known to display a  non-linear correlation of the form $\lr \propto \lx^{0.5-0.7}$ that extends unbroken over more than five orders of magnitude in $\lx$ \citep{corbel08, corbel13, gallo14}, which we refer to as the ``standard'' radio/X-ray correlation.   However, there is  a population of ``radio-faint'' \xrb s at $\lx \gtrsim 10^{36}~\ergs$ ($\gtrsim10^{-3} \ledd$) with radio luminosities that are 1-2 decades fainter than predicted by the ``standard'' correlation \citep[e.g.,][]{corbel04, cadolle-bel07, rodriguez07, xue07, soleri11, gallo12}, some of which show  correlations as steep as $\lr \propto \lx^{1.4}$ \citep{coriat11, cao14}.   Intriguingly, the ``radio-faint'' \xrb\ with the best radio/X-ray coverage, \hsev,  was unexpectedly observed to move horizontally across the $\lr-\lx$ plane when $\lx \lesssim 4 \times 10^{36}~\ergs$, until it rejoined the ``standard'' track around $\lx \approx 10^{35}~\ergs$ \citep{jonker10, coriat11}.  Two other systems, \maxijsixteen\ and \xtejseventeen, appeared to also take  similar paths between the two tracks \citep{jonker12, ratti12}.

The radio/X-ray luminosity plane becomes more poorly sampled as one moves toward lower luminosities \citep{miller-jones11}.  
The \xrb\ candidate \fullsrcname\ (hereafter \src) recently afforded an opportunity to improve our coverage  at low luminosities.  \src\ was discovered in outburst in 2005 \citep{palmer05}, where it surprisingly remained  for almost 12 years.    The end of the outburst was  noticed at optical wavelengths from 2016 September -- November  \citep{russell16} by a program that  regularly monitors $\sim$40 \xrb s with the Faulkes Telescope Project \citep{lewis08}.   \src\ then underwent a  mini-outburst\footnote{The optical flux peaked at a similar magnitude as before the initial descent into quiescence (see Zhang et al.\ in prep).  Following \citet{chen97}, we refer to such renewed activity as a mini-outburst.  We also note that \src\ was too close to the Sun to observe from mid-November 2016 through mid-January 2017 in the optical and X-ray wavebands.} 
  that was first detected in late January of 2017 and lasted through mid-April 2017 \citep[][]{bright17, kong17, shaw17, qasim17, tomsick17}.  In late April 2017, approximately a week after \src\ returned below radio/X-ray detection thresholds, it  underwent a second mini-outburst \citep{bernardini17}.  \src\ returned to quiescence in the optical waveband by  July 2017 \citep{zhang17}.

Here, we present results on the radio/X-ray luminosity correlation during  the mini-outburst(s) of \src, where we cover the radio/X-ray luminosity plane over the sparsely sampled $\lx \approx 10^{33}-10^{36}~\ergs$ regime.  One particularly useful property of \src\ for studying \xrb s at low luminosities is that \src\ lies at high Galactic latitude ($b= 12^\circ$).  Therefore, the line of sight absorption is relatively small ($\nh \approx 2.0\pm0.3 \times 10^{21}~\cmtwo$; \citealt{froning14}), allowing X-ray detections at lower luminosities than for most  \xrb s.  \src\ is also an intriguing target because it was established as a ``radio-faint'' \xrb\ \citep[e.g.,][]{cadolle-bel07, soleri10} where it followed $\lr \propto \lx^{0.96}$ \citep{rushton16} while in the hard state.  \src\ is  likely to host a black hole instead of a neutron star.  For example, from the width of the (disk) H$\alpha$   emission line in outburst, \citet{shaw16b} derived a compact object mass $\gtrsim 7 \msun$.  Furthermore,   during its outburst, the X-ray spectral and timing properties appeared  more similar to other \xrb s than to neutron star X-ray binaries \citep[e.g.,][]{cadolle-bel07, durant09, soleri10}.
The  distance to \src\ is suggested to fall between 2--8 kpc \citep{cadolle-bel07, froning14}.   Following \citet{rushton16},  we adopt 8 kpc here, although adopting a lower value does not (qualitatively) alter our conclusions.   Unless stated otherwise, we  define X-ray luminosities from 1-10 keV, and we report uncertainties on radio and X-ray parameters at the 68 and 90\% confidence levels, respectively.

\section{Observations}
\label{sec:obs}
Our dataset combines observations  from  the Arcminute Microkelvin Imager Large Array  (AMI-LA; AMI Consortium: \citealt{zwart08, hickish17}),  the Karl G. Jansky Very Large Array (VLA), and the  Very Long Baseline Array (VLBA) in the radio; and from  the X-ray Telescope \citep[XRT;][]{burrows05} on board the \textit{Swift} X-ray Mission \citep{gehrels04}.  Our analysis is described below and summarized in Table~\ref{tab:obslog}.

\subsection{AMI-LA}
\label{sec:obs:ami}
AMI-LA monitored \src\ starting on 2017 February 15, observing a total of 35 times over $\sim$100 days (although the last AMI-LA detection was on 2017 April 8, about midway through our campaign).  Observations generally lasted  3--4 hours, with typical image noises $\sigma_{\rm rms} \approx 0.04$ mJy beam$^{-1}$.  Observations were carried out at a central frequency of $15.5\,\mathrm{GHz}$ with a total bandwidth of $5\,\mathrm{GHz}$.  We observed the calibrator source J1804+0101 for $\sim$2 minutes for every 9 minutes on source  to find  the complex gain solutions.  Data were binned into 8 channels, each with a  width of $625\,\mathrm{MHz}$, and the data were calibrated and flagged for radio frequency interference (RFI) with the AMI reduction pipeline {\sc reduce\_dc}.   Further RFI flagging was performed in  the Common Astronomy Software Application {\sc casa} v4.2.2 \citep{mcmullin07} and  imaging was performed with the task {\sc clean}, setting a halting threshold of 3$\sigma_{\rm rms}$.  To extract flux measurements we used the {\sc python} based source extractor {\sc pyse}, which was developed as part of the LOFAR Transient Pipeline ({\sc trap}; \citealt{swinbank15}). A two dimensional Gaussian with the same dimensions as the synthesized beam  was used to fit sources in the image plane. We detected an unresolved source consistent with the location of \src\ in eight of our observations, using a detection threshold of $3.5\sigma_{\rm rms}$ and including all pixels with values $>$3$\sigma_{\rm rms}$ during the  fitting analysis, where $\sigma_{\rm rms}$ is the statistical error (the error bars reported in Table~\ref{tab:obslog} also include a 10\% systematic error from uncertainties on  the flux density calibration scale).    Throughout this paper we only consider these eight detections and ignore upper limits, since we initiated our (more sensitive) VLA observations shortly after \src\ was no longer detected by AMI-LA.

\subsection{VLA}
\label{sec:obs:vla}

We observed \src\ with the VLA after its initial descent into quiescence on 2016 November 5 and 7 (project code VLA/16A-060, see \citealt{plotkin16}), and we also obtained three epochs during the mini-outbursts on 2017 April 19, 21, and 29 (project code VLA/17A-430, awarded through Director''s Discretionary Time).  The observational setups were similar for all observations, except that the VLA was in the most extended (A) configuration during the 2016 November observations, and it was in the most compact (D) configuration during the observations from 2017.

We used  two basebands centered at 9.0 and 10.65 GHz, with 1.9 and 1.8 GHz bandwidth respectively.    Observations lasted for 1 hour each ($\approx$32--38 min on source), except for 2016 November 7 which lasted for 2.25 hours ($\approx$105 min on source).   We observed the phase calibrator J1743$-$0350 every 5--8 min to solve for the complex gains, and we set the flux amplitude scale using 3C~286 on 2016 November 5, 7 and 2017 April 19 and 3C~48 on the other two epochs.    Weather conditions were poor on 2016 November 5, and we could not obtain useful phase solutions to calibrate the data.  Weather was good during the other four epochs.  

Data were processed using standard procedures in  {\sc casa} v 4.7.1, and the flux scale was set using the task {\sc setjy} and the \citet{perley13} coefficients.
We  imaged the field with the task {\sc clean}, using two Taylor terms to model the frequency dependence of other sources in the field,  and Briggs weighting with robust=1 to reduce sidelobes from other sources in the field.    \src\ was detected on 2017 April 19 and April 29, with peak flux densities of 45 (6.1$\sigma_{\rm rms}$) and 19 (3.6$\sigma_{\rm rms}$) $\mu$Jy bm$^{-1}$ at 9.8 GHz, respectively, measured with the task {\sc imfit} by fitting a point source model in the image plane.  No radio emission was detected from \src\ on 2016 November 7 or on 2017 April 21, and we derived 3$\sigma_{\rm rms}$ upper limits of $f_{\nu} < 8$ $\mu$Jy bm$^{-1}$ and $f_{\nu} < 16$ $\mu$Jy bm$^{-1}$, respectively.  Error bars on VLA flux densities in Table~\ref{tab:obslog} include statistical errors and 5\% systematic errors (the latter is the accuracy on the VLA flux density calibration scale).

\begin{table}[htbp]
\begin{deluxetable*}{c c c C C C C C}
\tablecaption{Summary of Radio and X-ray Observations \label{tab:obslog}}
\decimals
\tablecolumns{8}
\renewcommand\arraystretch{1.2}
\tablehead{
                \colhead{Date}                               & 
                \colhead{Telescope}                       & 
                \colhead{MJD}                               & 
                \colhead{$f_r$}                               & 
                \colhead{$\log \left(\nu L_{\nu}\right)_{\rm 5\,GHz}$}             & 
                \colhead{$\Gamma$}                     & 
                \colhead{$f_{0.6-10\,{\rm keV}}$}                   & 
                \colhead{$\log L_{1-10\,{\rm keV}}$}    \\   
                \colhead{}                                      & 
                \colhead{}                                       & 
                \colhead{}                                      & 
                \colhead{(mJy bm$^{-1}$)}     & 
                \colhead{($\ergs$)}                        & 
                \colhead{}                                       & 
                \colhead{($10^{-12}\,\flux$)}           &    
                \colhead{($\ergs$)}                         
}
\colnumbers
\startdata               
2016 Nov  6\tablenotemark{a} & XRT/PC     &  57698.08592 &  \nodata              & \nodata              & 1.7\tablenotemark{c}                   & <0.3                 & <33.3                \\
2016 Nov  7\tablenotemark{a} & XRT/PC     &  57699.48564 &  \nodata              & \nodata              & 1.7\tablenotemark{c}                   & <0.2                 & <33.0                \\
2016 Nov  7\tablenotemark{b}  & VLA        &  57699.91875 &  <0.008               & <27.5                & \nodata              & \nodata              & \nodata              \\
2017 Feb 15 & AMI-LA     &  57799.23803 &  0.291 \pm 0.055      & 29.0 \pm 0.1         & \nodata              & \nodata              & \nodata              \\
2017 Feb 16 & XRT/PC     &  57800.07281 &  \nodata              & \nodata              & $1.8\pm0.2$          & 61.8^{+29.6}_{-19.5} & 35.6^{+0.2}_{-0.1}   \\
2017 Feb 19 & AMI-LA     &  57803.25274 &  0.346 \pm 0.048      & 29.1 \pm 0.1         & \nodata              & \nodata              & \nodata              \\
2017 Feb 19 & XRT/PC     &  57803.33760 &  \nodata              & \nodata              & $1.9\pm0.2$          & 71.4^{+45.8}_{-23.5} & 35.7^{+0.3}_{-0.1}   \\
2017 Feb 22 & XRT/PC     &  57806.39433 &  \nodata              & \nodata              & $1.6\pm0.1$          & 81.7^{+22.0}_{-13.8} & 35.7^{+0.1}_{-0.1}   \\
2017 Feb 23 & XRT/WT     &  57807.91082 &  \nodata              & \nodata              & $1.8\pm0.1$          & 82.7^{+23.2}_{-13.8} & 35.7^{+0.1}_{-0.1}   \\
2017 Feb 24 & XRT/WT     &  57808.91153 &  \nodata              & \nodata              & $1.8\pm0.2$          & 98.9^{+36.0}_{-23.3} & 35.8^{+0.2}_{-0.1}   \\
2017 Feb 25 & XRT/PC     &  57809.31692 &  \nodata              & \nodata              & $1.4\pm0.2$          & 97.3^{+60.4}_{-33.2} & 35.8^{+0.3}_{-0.1}   \\
2017 Mar 14 & AMI-LA     &  57826.23219 &  0.223 \pm 0.048      & 28.9 \pm 0.1         & \nodata              & \nodata              & \nodata              \\
2017 Mar 22 & AMI-LA     &  57834.13557 &  0.283 \pm 0.051      & 29.0 \pm 0.1         & \nodata              & \nodata              & \nodata              \\
2017 Mar 25 & AMI-LA     &  57837.11836 &  0.173 \pm 0.050      & 28.8 \pm 0.1         & \nodata              & \nodata              & \nodata              \\
2017 Mar 25 & XRT/WT     &  57837.61163 &  \nodata              & \nodata              & $1.7\pm0.2$          & 38.6^{+25.2}_{-8.6}  & 35.4^{+0.3}_{-0.1}   \\
2017 Mar 27 & AMI-LA     &  57839.10113 &  0.199 \pm 0.044      & 28.9 \pm 0.1         & \nodata              & \nodata              & \nodata              \\
2017 Apr  1 & XRT/WT     &  57844.01194 &  \nodata              & \nodata              & $2.2\pm0.3$          & 48.6^{+102.6}_{-31.2} & 35.5^{+0.9}_{-0.3}   \\
2017 Apr  1 & AMI-LA     &  57844.09925 &  0.201 \pm 0.044      & 28.9 \pm 0.1         & \nodata              & \nodata              & \nodata              \\
2017 Apr  6 & XRT/WT     &  57849.58618 &  \nodata              & \nodata              & $1.8\pm0.3$          & 22.4^{+16.7}_{-6.8}  & 35.2^{+0.3}_{-0.1}   \\
2017 Apr  8 & AMI-LA     &  57851.10923 &  0.202 \pm 0.056      & 28.9 \pm 0.1         & \nodata              & \nodata              & \nodata              \\
2017 Apr  8 & XRT/WT     &  57851.24913 &  \nodata              & \nodata              & $2.3\pm0.4$          & 23.1^{+17.8}_{-7.4}  & 35.1^{+0.3}_{-0.1}   \\
2017 Apr 13 & VLBA       &  57856.54167 &  <0.160               & <28.8                & \nodata              & \nodata              & \nodata              \\
2017 Apr 15 & XRT/PC     &  57858.01942 &  \nodata              & \nodata              & $1.8\pm0.6$          & 4.9^{+8.5}_{-2.3}    & 34.5^{+0.7}_{-0.2}   \\
2017 Apr 18 & XRT/PC     &  57861.87166 &  \nodata              & \nodata              & $1.5\pm0.5$          & 1.9^{+2.4}_{-1.4}    & 34.1^{+0.5}_{-0.3}   \\
2017 Apr 19 & VLA        &  57862.30750 &  0.045 \pm 0.007      & 28.2 \pm 0.1         & \nodata              & \nodata              & \nodata              \\
2017 Apr 20 & XRT/WT     &  57863.98851 &  \nodata              & \nodata              & 1.7\tablenotemark{c}                  & <2.0                 & <34.1                \\
2017 Apr 21 & VLA        &  57864.60620 &  <0.016               & <27.8                & \nodata              & \nodata              & \nodata              \\
2017 Apr 22 & XRT/PC     &  57865.91086 &  \nodata              & \nodata              & 1.7\tablenotemark{c}                  & <0.2                 & <33.1                \\
2017 Apr 29 & VLA        &  57872.59650 &  0.019 \pm 0.005      & 27.9 \pm 0.1         & \nodata              & \nodata              & \nodata              \\
2017 Apr 29 & XRT/PC     &  57872.77287 &  \nodata              & \nodata              & 1.7\tablenotemark{c}                  & 0.2^{+0.4}_{-0.2}    & 33.2^{+0.8}_{-0.3}   \\
2017 May  6 & XRT/PC     &  57879.67537 &  \nodata              & \nodata              & $1.8\pm0.4$          & 2.4^{+4.4}_{-1.3}    & 34.2^{+0.8}_{-0.2}   \\
2017 May 15 & XRT/PC     &  57888.43380 &  \nodata              & \nodata              & 1.7\tablenotemark{c}                  & 0.2^{+0.4}_{-0.2}    & 33.2^{+0.8}_{-0.3}   \\
 \enddata
\vspace{0.3cm}
\tablenotetext{a}{Observation first reported by \citet{shaw16}.}
\tablenotetext{b}{Observation first reported by \citet{plotkin16}.}
\tablenotetext{c}{Due to a low number of photons, X-ray fluxes were estimated using an absorbed power-law model with photon index $\Gamma=1.7$  and a column density $\nh=2\times10^{21}~\cmtwo$. }
\tablecomments{Column (1) calendar date of each observation.  
Column (2) the telescope used for each observation.  For \textit{Swift}/XRT, we specify if the observations were taken in photon counting (PC) or window timing (WT) mode.
Column (3) modified julian date of each observation.
Column (4) peak radio flux density at the central observing frequency (15.5 GHz for AMI-LA , 9.8 GHz for the VLA, and 4.98 GHz for the VLBA).  All radio error bars are reported at the 68\% confidence level (and they include systematic errors on the flux density calibration scale) and upper limits are at the 3$\sigma_{\rm rms}$ level for the VLA and 5$\sigma_{\rm rms}$ for the VLBA.
Column (5) logarithm of the radio luminosity at 5 GHz, assuming a flat radio spectrum (see Section~\ref{sec:res:lrlx}) and $d=8\, {\rm kpc}$.
Column (6) best-fit photon index $\Gamma$ from each \textit{Swift} observation.   All X-ray error bars are reported at the 90\% confidence level.  
Column (7) model X-ray flux over the  0.6-10 keV \textit{Swift}/XRT energy band.  Upper limits are at the 99\% confidence level.
Column (8) logarithm of the   1-10 keV X-ray luminosity, assuming $d=8\,{\rm kpc}$.
}
\end{deluxetable*}
\end{table}
\renewcommand\arraystretch{1}

\subsection{VLBA}
\label{sec:obs:vlbi}
We observed \src\ with the VLBA on 2017 April 13 (10:30--15:30 UT), as part of a filler-time astrometric program (project code BM449).  We observed at a central frequency of 4.98 GHz, with an observing bandwidth of 256\,MHz.  30\,min at the start and end of the observation was dedicated to a geodetic block, observing a range of bright calibrators across the sky to correct for unmodeled tropospheric and clock errors in the correlated data.  For the remaining four hours, we switched between phase reference calibrator sources, \src, and an astrometric check source.  We used the nearby compact source J1752$-$0147 (RA=17$^{\rm h}$52$^{\rm m}$18$^{\rm s}$.3637813, Dec=$-$01$^\circ$47$\arcmin$16.685462$\arcsec$ (J2000); only 27\,arcmin from \src) as our primary phase reference calibrator, using a 3-minute cycle time (110\,s on target, 70s on calibrator), and we observed the brighter but more distant calibrator J1743$-$0350 every 20\,min to calibrate the delays and rates.  The data were calibrated according to standard procedures within the Astronomical Image Processing System \citep[{\sc aips};][]{greisen03}.  No radio source was detected above a 5$\sigma_{\rm rms}$ upper limit of 0.16\,mJy\,bm$^{-1}$ (the systematic uncertainty on the amplitude calibration is  $\approx$5\%).

\subsection{Swift XRT}
\label{sec:obs:xrt}

X-ray observations were taken with \textit{Swift}/XRT shortly after the initial outburst decay (on 2016 November 6 and 7; see \citealt{shaw16} for details), and during the mini-outbursts from 2017 February 16 -- 2017 May 15 (Target ID: 00030090). \emph{Swift}/XRT observed the source in auto-exposure mode for the majority of the observations, adjusting the CCD readout mode between windowed timing (WT) and photon counting (PC) according to the observed count rate.

Data were reprocessed using the \textsc{heasoft} v6.19\footnote{\url{https://heasarc.nasa.gov/lheasoft/}} task \textsc{xrtpipeline}. WT count rates were extracted using a circular region 20 pixels in radius ($\approx 47\arcsec$). Background count rates in WT mode were extracted from an annulus centered on the source with inner and outer radii of 80 and 120 pixels, respectively. PC mode source count rates were extracted from a circular region of the same radius as in WT mode, and the average count rate was then calculated in order to determine if photon pile-up was significant. PC observations with count rates higher than 0.5 counts s$^{-1}$ were re-extracted using an annulus with a 20 pixel outer radius and the central portion of the point spread function excluded. The radius of the excluded region was determined using NASA''s \textsc{ximage} package\footnote{\url{http://www.swift.ac.uk/analysis/xrt/pileup.php}} and ranged from $\sim$2-4 pixels. PC mode background count rates were extracted from an annulus centered on the source with inner and outer radii of 50 and 70 pixels, respectively.   The number of (net) source counts ranged from $\approx$10-2500 counts.

Spectra of each observation were extracted and spectral fits were performed in {\sc xspec} v12.9.0 \citep{arnaud96}.  Due to the small number of counts in some observations (8 epochs have $<$100 net counts), we grouped each spectrum to have a minimum of one count per energy bin, and we performed the spectral fitting using  Cash statistics for background subtracted spectra \citep[W-statistics;][]{cash79}.  Interstellar absorption was accounted for by the \textsc{tbabs} model with \citet{wilms00} abundances and \citet{verner96} photoionization cross-sections.  We obtained adequate fits to all spectra using an absorbed power-law model (\textsc{powerlaw}; no model fit was improved by adding a  \textsc{diskbb} component).  We then extracted unabsorbed model fluxes and 90\% error bars in the 0.6-10 keV and the 1-10 keV bands with the tool {\sc cflux} (the error bars also incorporate uncertainties related to the best-fit model parameters).

For observations without enough X-ray counts to fit a spectral model ($\lesssim$30-50 counts), we required detections to be significant at the $>$99\% confidence level, according to Poisson statistics in the presence of background \citep{kraft91}.  For these low-count observations, we assumed a power-law model with $\Gamma=1.7$ and  $\nh=2\times10^{21}$ cm$^{-2}$ \citep{froning14} to estimate a flux.  For error bars we adopted 90\% confidence intervals from \citet{kraft91}, and  we factored in a photon index  that was allowed vary from $1<\Gamma<2.5$.

\section{Results}
\label{sec:res}

\subsection{Light Curves}
\begin{figure}
	\includegraphics[scale=0.57]{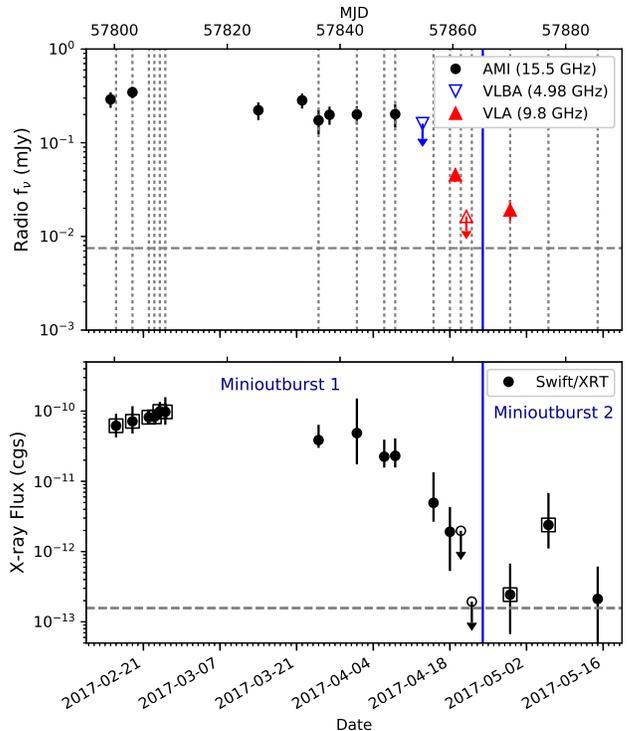}
    \caption{Radio and X-ray light curves of the 2017 mini-outburst(s).  The top panel shows radio flux densities from AMI-LA (15.5 GHz; black circles), the VLBA (4.98 GHz; blue upside down triangle), and the VLA (9.8 GHz; red triangles), with error bars representing 68\% confidence.  The vertical dotted lines mark the epochs of our X-ray observations.  The bottom panel  shows X-ray fluxes (in units of $\flux$) from 0.6-10 keV with errors illustrated at 90\% confidence.   The vertical blue solid line illustrates the approximate boundary between the two mini-outbursts, and the data points circumscribed by squares represent when \src\ was rising out of quiescence.  The horizontal dashed lines show the deepest radio and X-ray flux limits yet for \src\ in quiescence,  both obtained on 2016 November 7. }
      \label{fig:lc}
\end{figure}

Radio  and X-ray  light curves  are displayed in Figure~\ref{fig:lc}, which span from  February through May of 2017.  \src\ dropped below our radio and X-ray detection thresholds from April 20-22 (despite being detected at both wavebands two days earlier).  \src\  was subsequently  detected at both wavebands again on April 29, implying a second mini-outburst.  During the rise of the second mini-outburst, we caught \src\ shortly after it brightened above our detection thresholds.  

\begin{figure*}
\begin{center}
	\includegraphics[scale=0.75]{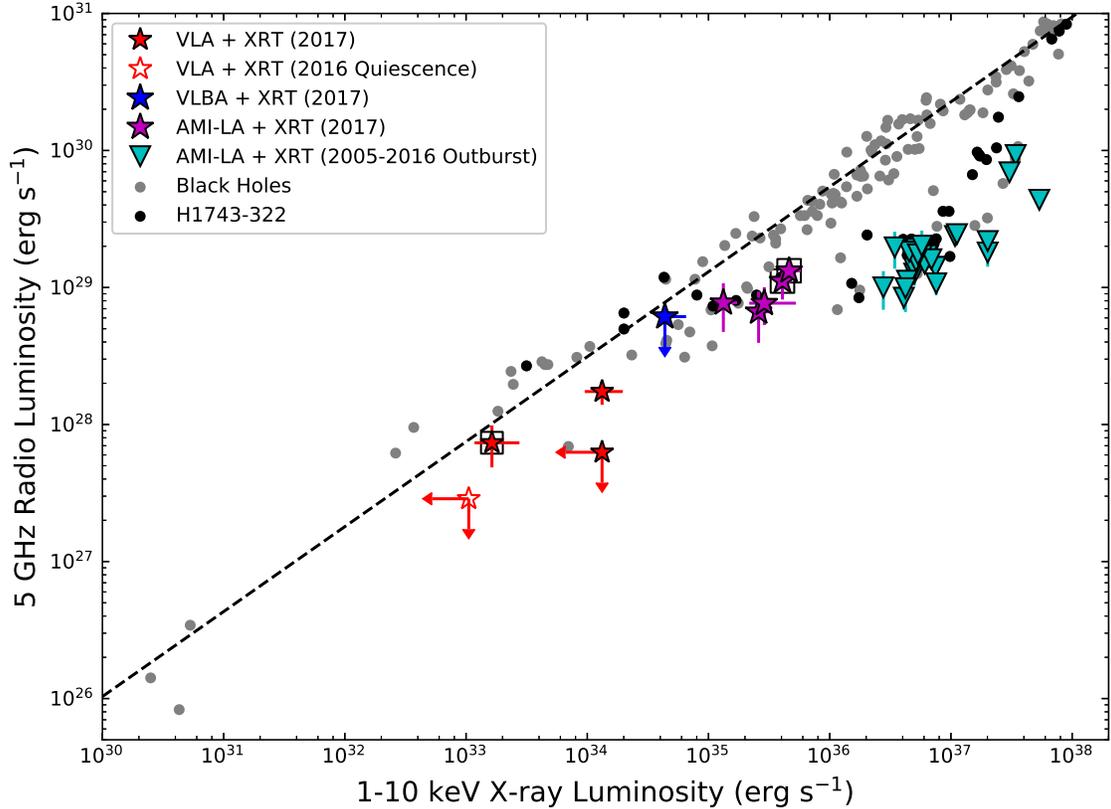}
    \caption{\src\ on the  radio/X-ray luminosity plane,  assuming it is located 8 kpc away.  The red star  symbols show epochs with the VLA (the open symbol represents our deeper limit in quiescence from 2016 November), the blue star symbol with the VLBA, and the purple star symbols  with AMI-LA.  Data points during the rise out of quiescence (three epochs) are circumscribed by  squares.  The cyan upside down triangles show the location of \src\ in the ``radio-faint'' hard state during its $\gtrsim$11 year outburst (data taken from \citealt{rushton16}).  For uniformity, all X-ray error bars are rescaled to 68\% confidence.   A sample of comparison \xrb s are shown as grey circles, with \hsev\ highlighted with  dark circles.  The dashed line shows the radio/X-ray correlation of \gxthree\ from \citet{gallo14}, to illustrate the ``standard'' track (see Section \ref{sec:res:lrlx}).  \src\ lies close to the ``standard'' track at low luminosities $\lx \lesssim 10^{36}~\ergs$.  } 
       \label{fig:lrlx}
       \end{center}
\end{figure*}

\subsection{Radio/X-ray Correlation}
\label{sec:res:lrlx}

A total of eight of our radio (AMI-LA/VLA) and X-ray observations were taken $<$1 day apart, which we place on the radio/X-ray luminosity plane in Figure~\ref{fig:lrlx}.  Our VLBA radio limit from 2017 April 13  was taken 1.5 days before an X-ray observation.  To place that epoch on the radio/X-ray plane, we interpolate the X-ray light curve between 2017 April 8--18 to the time of the VLBA observation (the X-ray light curve appears to be exponentially decaying during that time period; see Figure~\ref{fig:lc}).  We  attempt to improve the ``simultaneity'' of the other eight epochs by interpolating the radio and X-ray light curves, but doing so does not  alter any results (within  errors). Our observations include  epochs when the X-ray flux is rising (three data points circumscribed by squares in Figure~\ref{fig:lrlx}) and others when \src\ is fading back into quiescence, providing a rare opportunity to compare disk/jet couplings in both directions. 

 For ease of comparison to the literature  we extrapolate all radio observations to 5 GHz, assuming a flat radio spectrum.   The assumption of a flat radio spectrum appears reasonable from \citet{cadolle-bel07}, who measured $\alpha_r = 0.03 \pm 0.03$ ($f_\nu \propto \nu^{\alpha_r}$) for \src\ from radio observations in 2005.  However, \citet{tomsick15} measured an inverted radio spectrum from  observations taken in 2014 ($\alpha_{\rm r} = 0.29 \pm 0.05$).  We therefore add  uncertainties to the radio luminosity error bars in Figure~\ref{fig:lrlx} to account for a radio spectrum that could be as inverted as $\alpha_{\rm r} = 0.3$.  

    For comparison, we also display  in Figure~\ref{fig:lrlx} radio and X-ray observations of \src\ during its 2005-2016 outburst, when \src\ was in the radio-faint hard state \citep{rushton16}.  To illustrate the ``standard'' track in Figure~\ref{fig:lrlx}, we adopt the best-fit to the \xrb\ \gxthree\ from \citet{gallo14} (we use \gxthree\ as a representative example because it has the most data coverage for any ``standard'' track \xrb, taken over multiple outbursts; \citealt{corbel13}).  We also highlight the path  \hsev\  took through the $\lr-\lx$ plane to emphasize that \src\ appears to occupy a similar parameter space.

\section{Discussion}
\label{sec:disc}
Figure~\ref{fig:lrlx} shows that during the 2017 mini-outbursts \src\ occupies a region of the $\lr-\lx$ plane that is inconsistent with  the $\lr \propto \lx^{0.96}$ correlation it followed during its outburst \citep{rushton16}.    Even though \src\ appears to always fall below the radio/X-ray correlation defined by \gxthree\ (dashed solid line),  its path through the radio/X-ray plane is clearly different above and below $\lx \approx 10^{36}~\ergs$ (i.e., \src\ does not simply follow a single, lower-normalization correlation that is parallel to the ``standard track'').  However, it is unclear  whether the difference above and below $10^{36}~\ergs$ is driven by the X-ray luminosity, by the Eddington ratio, or by other details related to the physics of the mini-outbursts.
  
Intriguingly, \src\ occupies a similar parameter space in Figure~\ref{fig:lrlx} as \hsev\ (black circles), a \xrb\ that was observed to move horizontally across the $\lr-\lx$ plane as it transitioned from the hard state to quiescence at the end of an outburst \citep{jonker10, coriat11}.  It very likely could be the case that \src\ also moved horizontally across the $\lr-\lx$ plane.  However, since the initial decay was not monitored in the radio and X-ray in 2016, we cannot exclude a scenario where \src\ faded down its ``radio-faint'' hard state $\lr \propto \lx^{0.96}$ correlation during the initial decay, and  then it rose and faded along a path close to the ``standard'' track during the mini-outbursts.  Unfortunately, \src\ did not reach high enough X-ray luminosities during the mini-outbursts to determine if it would have moved horizontally back to the ``radio-faint'' hard state.   Regardless, our campaign reinforces the notion that ``radio-faint'' \xrb s  can follow paths similar to the ``standard'' track at the lowest X-ray luminosities, and we still lack  observational evidence for the existence of a ``radio-faint'' \xrb\ branch below $\lx \approx 10^{36}~\ergs$. 
 
We note that while \src\ is illustrated in Figure~\ref{fig:lrlx} assuming a distance of 8 kpc,  it has also been suggested that the source distance could be as low as 2--4 kpc \citep[e.g.,][]{cadolle-bel07, froning14}.  Adopting a  lower distance would of course not change our primary conclusion that \src\ appears to follow different radio/X-ray correlations at high and low luminosities.  However, if \src\ were to be  closer than 8 kpc, then it would fall even farther below the ``standard'' track at low luminosities, and it would not occupy precisely the same parameter space as \hsev.  The better agreement with other \xrb s on $\lr-\lx$ at 8 kpc might suggest that \src\ indeed lies at a larger distance.   The distance estimate will hopefully be improved in the future through   studies on the quiescent optical counterpart.

\subsection{The radio/X-ray correlation in quiescence}
\label{sec:disc:quiesc}
As \xrb s fade toward quiescence they may enter a jet-dominated state, where a substantial fraction of the accretion power could be channeled into the jet as mechanical power instead of being liberated as X-rays from within the accretion flow \citep{fender03}.\footnote{Some of the accretion power can also be advected through the black hole event horizon \citep[e.g.,][]{garcia01}.}  
  \citet{yuan05} predict that at luminosities below $\lx \approx 10^{33}-10^{34}~\ergs$ ($10^{-6}-10^{-5}~\ledd$) that the jet may also dominate the radiative output, with the observed X-ray emission arising predominantly from non-thermal emission from a synchrotron cooled jet.   As a consequence, \citet{yuan05} predict that the radio/X-ray luminosity correlation  will follow a steeper slope in quiescence.    Our campaign on \src\  detected radio and X-ray emission near the ``standard'' track at a luminosity as low as $\lx \approx 2 \times 10^{33}~\ergs$ ($\approx10^{-5.7}~\ledd$),  implying that if the ``standard'' track steepens, then it must do so at an even lower luminosity.  \citet{plotkin17} more rigorously showed for the \xrb\ \vfour\ that the ``standard'' radio/X-ray correlation maintains its slope  to at least   $\lx \approx 3 \times 10^{32}~\ergs$ ($\approx10^{-6.5}\ledd$) in that source.  Furthermore,  radio detections of the \xrb s \asix, \mwcsix, and \xtejeleven\ all fall on an extrapolation of the  ``standard'' track to $\lx \approx 2\times10^{30} - 1\times10^{31}~\ergs$ ($\approx 10^{-8.7} - 10^{-8.0}~\ledd$; \citealt{gallo06,gallo14, ribo17}).  From the above, it seems reasonable to  exclude the possibility that all \xrb s follow a steeper radio/X-ray correlation at the lowest luminosities. 

By incorporating a mass normalization term, the ``standard'' radio/X-ray correlation can be extended to include supermassive black holes that power low-luminosity active galactic nuclei (LLAGN); i.e.,   the fundamental plane of black hole activity \citep{merloni03, falcke04}.        Intriguingly, \citet{xie17} find a  steepening of the slope of the fundamental plane for  quiescent LLAGN at $\lx \lesssim 10^{-6}~\ledd$ \citep[also see][]{yuan09}, in line with the predictions of \citet[][]{yuan05} (although see \citealt{dong15} for an alternative view). There is thus some tension toward understanding why a steeper correlation may exist for quiescent LLAGN, while  observations of \xrb s so far do not show any evidence for a steepening.  Whether or not there is a single ``track'' in quiescence will have  consequences not just on our understanding of quiescent accretion and jet physics, but also on our ability to use the fundamental plane to search for new populations of quiescent \xrb s \citep[e.g.,][]{maccarone05, strader12, chomiuk13, fender13, miller-jones15, tetarenko16} and LLAGN, particularly in the intermediate mass range \citep[e.g.,][]{miller-jones12, cseh15, koliopanos17, mezcua17}.

  One explanation for the apparent difference in the radio/X-ray correlation slope between quiescent \xrb s and LLAGN is that we simply have not yet observed enough \xrb s to detect a (sub)population that proceeds down a steeper track in quiescence.  It is also possible that for LLAGN an extra source of X-ray emission could be contributed by X-ray binaries near the nucleus of the host galaxy \citep[e.g.,][]{miller15}, which could artificially steepen the radio/X-ray correlation slope (and add additional scatter).  A third explanation, which we explore in more detail below,  is that \xrb s do not have black holes that are massive enough for a synchrotron cooled jet to ever dominate the X-ray waveband. 
  
For a non-thermal distribution of synchrotron emitting particles accelerated along a jet, the frequency above which particles suffer synchrotron radiative cooling losses scales as $\nu_c \propto \mdot^{-3/2} \mbh^{-1/2}$ \citep{heinz04}, where $\mdot$ is the Eddington normalized mass accretion rate ($\Mdot/\Mdot_{\rm Edd}$).\footnote{In general, $\lx/\ledd$ can be used as a rough proxy for $\mdot$, but we stress that $\lx/\ledd \neq \mdot$.}  
Thus, scaling from stellar mass ($\approx 10 \msun$) to supermassive scales ($\approx 10^6-10^9  \msun$) would lower the frequency of the synchrotron cooling break by $\approx$2.5-4 decades in frequency.   It could therefore be possible for synchrotron cooled radiation to appear in the X-ray waveband for supermassive black holes but not for \xrb s, purely from mass-scaling arguments that are independent of accretion rate (see \citealt{plotkin12} for observations supporting this interpretation). If this is correct, then jet emission can only dominate the X-ray spectrum of quiescent \xrb s if it is synchrotron self-Compton, and/or if the jet is not radiatively cooled\footnote{An uncooled jet would require less efficient particle acceleration, in order to be consistent with the typically soft X-ray spectra  ($\Gamma \approx 2$) of quiescent \xrb s; \citep[e.g.,][]{plotkin13, reynolds14}.} 
  \citep{gallo07, corbel08, plotkin15, plotkin17, connors17}; otherwise, the radiatively inefficient accretion flow will always dominate the X-ray spectrum of quiescent \xrb s. \citep[e.g.,][]{esin97, mcclintock03, zdziarski04, sobolewska11, qiao13, yuan14}.

\subsection{Comparing the rise and decay out of quiescence}
\label{sec:disc:direction}

\begin{figure*}
\begin{center}
	\includegraphics[scale=0.7]{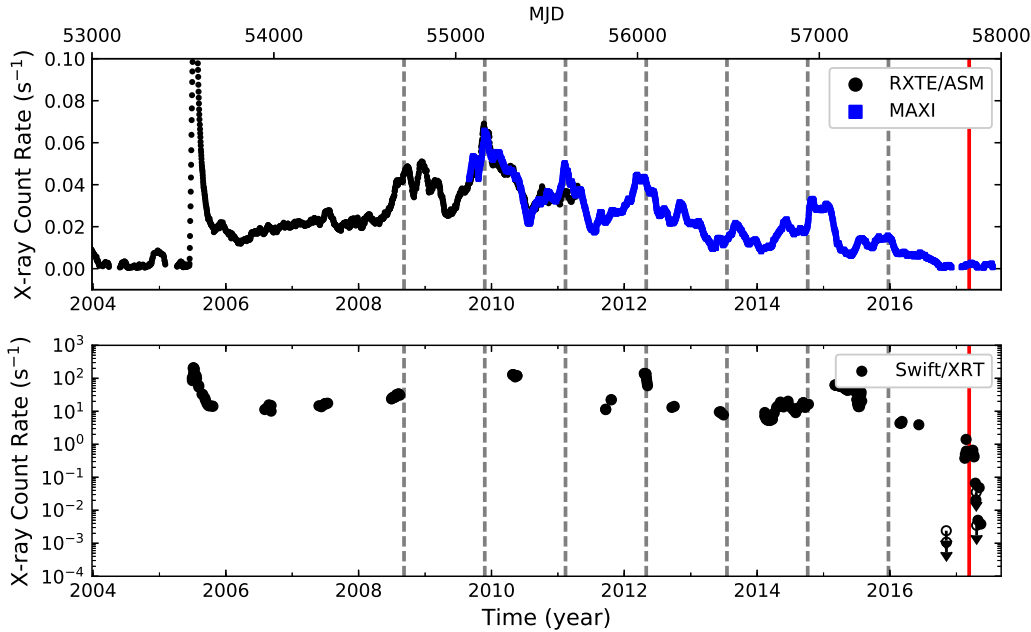}
    \caption{\textit{Top panel:} \textit{RXTE}/ASM (black circles) and \textit{MAXI} (blue squares) X-ray light curves (2-10 keV) toward \src, showing 30-day running averages from 2004-2017.  \textit{Bottom panel:} X-ray light curve from \textit{Swift}/XRT (0.3-10 keV).  The dashed grey lines are separated by 444 days, to mark long-term modulations (anchored to  two sharp peaks around MJD 55600 = 2011 February 8 and around 56050 = 2012 May 3).  The final vertical line is shaded red, to highlight that it coincides with the first 2017 mini-outburst.}
      \label{fig:longlc}
\end{center}
\end{figure*}

Radio and X-ray detections of a \xrb\ during the rise out of quiescence are rare at  low luminosities, and our radio/X-ray detections of \src\ at ($\lx, \lr) \approx (2\times 10^{33}, 8\times10^{27})~\ergs$ represent the lowest quasi-simultaneous luminosity detections yet in the rising hard state (for this data point, the radio/X-ray observations were only separated by 4.2 h). Even  the \xrb\ \gxthree, which has radio/X-ray coverage during its rise out of quiescence over multiple outbursts, only has (rising) radio/X-ray detections when $\lx \gtrsim 8 \times 10^{34}~\ergs$ (3-9 keV; \citealt{corbel13}, assuming a distance of $\sim$8 kpc).

There have been (tentative) suggestions that the direction in which a \xrb\ is moving can influence the normalization of the radio/X-ray correlation, and/or the high-energy radiation mechanisms.  For example, for \gxthree, the radio luminosity could be a factor of $\sim$two brighter during the hard state decay compared to during the rise \citep{corbel13}.  Also, \citet{russell10} find that hard X-rays (3-9 keV) from \xtejfifteen\  could be dominated by synchrotron jet emission during its hard state decay, while jet synchrotron only contributes up to a few percent during the rising hard state.\footnote{Some differences could also be luminosity dependent, since  observations of the rising hard state tend to probe higher X-ray luminosities than the decay, due  to the hysteretical behavior of \xrb\ outbursts \citep{maccarone03}.}

For \src, we do not observe a meaningful difference in the radio luminosities (relative to the X-ray) during the rising and decaying hard states.  The number of data points  is too small to fit for a correlation slope.  However, comparing to the radio/X-ray correlation for \gxthree\ \citep[$\lr \propto \lx^{0.62}$;][]{gallo14},  \src\ can fall 0.1-0.4 dex below the ``standard'' track in radio luminosity during the rise and 0.3-0.5 dex below the ``standard'' track during the decay, and we have no reason to suspect that the normalization of the radio/X-ray luminosity correlation is systematically different depending on the  direction.   In the hard state, the  jet  appears to respond to changes in the inner regions of the accretion flow/jet  on short timescales ($\lesssim$1--2 days), without a ``memory'' of whether the jet was brighter or fainter days earlier.   Such behavior might be expected: during soft-to-hard state transitions when the jet reactivates after being quenched in the thermal soft state, the radio jet usually appears to turn on at a time that nearly coincides with when the X-ray spectrum again becomes hard and non-thermal, even though it can take 10-30 days for the jet to brighten and become powerful  in the infrared waveband as the particle acceleration zone along the jet moves closer to the black hole \citep{miller-jones12, corbel13a, kalemci13, russell14}.   The negligible delay between the X-ray and radio wavebands could suggest that the response of the  jet to changing amounts of injected  power     operates on timescales comparable to the time it takes for material to travel  outward from the jet base (likely tens of minutes, based on causality arguments and a limited number of $\lesssim$10$^2$ au size constraints on hard state and quiescent \xrb s, e.g., \citealt{dhawan00, stirling01, miller-jones08, reid11, reid14, russell15, plotkin17}).

\subsection{Long-term flux modulations}
During the 2005 outburst of \src, \citet{shaw13} reported X-ray and optical modulations in the long-term light curve with a $\sim$420 day period.  In Figure~\ref{fig:longlc} (upper panel) we produce a light curve of \src\ in the 2--10 keV band over 30-day running averages over the entire 11-12 year outburst, using public data from the All-Sky Monitor onboard the {\it Rossi X-ray Timing Explorer} \citep[{\it RXTE}/ASM;][]{levine96} and  from the \textit{Monitor of All-sky X-ray Image} \citep[{\it MAXI};][]{matsuoka09}.  Data with large error bars ($>$2 cts/s for the ASM and $>$0.05 cts/s for \textit{MAXI}) were filtered out.  ASM and \textit{MAXI} count rates were divided by factors of 70 and 3, respectively, to normalize both to Crab units.

Long-term modulations can be seen in Figure~\ref{fig:longlc}, which we note are not strictly periodic.  The vertical lines  in Figure~\ref{fig:longlc} represent the expected peaks of a modulation with a (slightly longer) period of 444 days, which is anchored to the two sharp peaks around MJD 55600 (2011 February 8) and around 56050 (2012 May 3). Most of the vertical lines fall close to local maxima in the light curve. Interestingly, the last vertical line (around MJD 57820 = 2017 March 8) falls close to the peak of the first mini-outburst discussed in this work. To better demonstrate this, we show the long-term 0.6--10 keV   {\it Swift}/XRT light curve in the bottom panel of Figure~\ref{fig:longlc} (this light curve was assembled via the online \textit{Swift}/XRT data products generator; \citealt{evans09}). The fact that  the peak of the first mini-outburst discussed here is close in time to an expected maximum from the long-term modulations suggests that the  mini-outbursts are likely still part of the 11-12 year outburst event.  This interpretation is supported by  optical quiescence not being  reached until 2017 July \citep{zhang17}.

\subsection{Future Prospects}
\label{sec:conc}

Our campaign on \src\  demonstrates the feasibility of obtaining useful radio and X-ray detections in the sparsely sampled low-luminosity regime. 
The high Galactic latitude of \src\ is a major reason we are able to obtain X-ray detections at low X-ray luminosities approaching $10^{33}~\ergs$ with \textit{Swift}, while even more sensitive X-ray telescopes (e.g., \textit{Chandra} and \textit{XMM-Newton}) are required for most other systems.  Another reason why the $\lr-\lx$ is not well sampled at low luminosities is that, at these low-luminosities, most systems have X-ray fluxes below the detection thresholds of current X-ray all sky monitors.  It is therefore often difficult  to trigger  more sensitive X-ray observations until the source already has a luminosity above the low-luminosity regime of interest.   Our 2017 campaign on \src\ was  triggered from changes in its \textit{optical} flux (via regular monitoring with the Faulkes telescopes; Zhang et al.\ in prep), which resulted in VLA and \textit{Swift} detections at low luminosities.   Optical monitoring has  previously been shown to be a promising avenue for triggering (and interpreting) multiwavelength observations of \xrb s \citep[e.g.,][]{orosz97, jain01a, jain01b, bernardini16, russell17}.  Our campaign on \src\ further illustrates the utility of optical monitoring to improve  coverage of the low-luminosity accretion regime, allowing us to fill in a crucial parameter space to learn about how relativistic jets are coupled to their underlying accretion flows.

\section*{Acknowledgements}
We are grateful to the anonymous referee for helpful comments that improved this paper. We thank the VLA  for approving our request for DDT observations,  and Heidi Medlin and the VLA scientists for their help preparing our observations with a quick turnaround;  %
we thank Brad Cenko
for approving the \textit{Swift} TOO request and the \textit{Swift} duty scientists for carrying out
the observations; %
and we thank the staff at the Mullard Radio Astronomy Observatory for scheduling and carrying out the AMI-LA observations.   
We are grateful to Tony Rushton for providing radio and X-ray luminosities for \src\ in the radio-faint hard state, and to Elena Gallo and Gemma Anderson for helpful discussions.  
  The National Radio Astronomy Observatory is a facility of the National Science Foundation operated under cooperative agreement by Associated Universities, Inc.
 The AMI telescope is supported by the European Research Council under grant ERC-2012-StG-307215 LODESTONE, and the University of Cambridge.    We are grateful 
for IT knowledge exchange with the SKA project.   
This work made use of data supplied by the UK Swift Science Data Centre at the University of Leicester.  
The Faulkes Telescopes are maintained and operated by the Las Cumbres Observatory (LCO).  
RMP acknowledges support from Curtin University through the Peter Curran Memorial Fellowship.   
 JCAMJ is supported by an Australian Research Council Future Fellowship (FT140101082).  
  AWS is supported by an NSERC Discovery Grant and a Discovery Accelerator Supplement.  
  TDR acknowledges support from the Netherlands Organisation for Scientific Research (NWO) Veni Fellowship, grant number 639.041.646.  
AMMS \& TMC gratefully acknowledge support from the European Research Council under grant ERC-2012-StG-307215 LODESTONE.  
YCP acknowledges support from a Trinity College JRF.  


\bibliographystyle{apj}

\end{document}